%% file: main.tex
\documentclass[runningheads]{llncs}
\usepackage{graphicx}
\usepackage{booktabs}
\usepackage{hyperref}
\usepackage{amsfonts}
\usepackage{todonotes}
\usepackage{tabularx}

\usepackage{enumitem}

\newcommand{\query}{\mathbf{q}}
\newcommand{\doc}{\mathbf{d}}
\newcommand{\rel}{r}
\newcommand{\tuple}[1]{\langle#1\rangle}
\newcommand{\sourcecol}{\mathcal{S}}
\newcommand{\targetcol}{\mathcal{T}}

\begin{document}
\title{Teaching a New Dog Old Tricks: Resurrecting Multilingual Retrieval Using Zero-shot Learning}
\titlerunning{Multilingual Retrieval Using Zero-shot Learning}

\author{
Sean MacAvaney\inst{1} \and
Luca Soldaini\inst{2} \and
Nazli Goharian\inst{1}}

\institute{IR Lab, Georgetown University, Washington DC 20057, USA \email{\{firstname\}@ir.cs.georgetown.edu} \and
{Amazon Alexa Search, Manhattan Beach CA, 90266, USA \\
\email{lssoldai@amazon.com}} \\
}
\authorrunning{S. MacAvaney et al.}

\maketitle
\begin{abstract}
While billions of non-English speaking users rely on search engines every day, the problem of ad-hoc information retrieval is rarely studied for non-English languages. 
This is primarily due to a lack of data set that are suitable to train ranking algorithms. 
In this paper, we tackle the lack of data by leveraging pre-trained multilingual language models to transfer a retrieval system trained on English collections to non-English queries and documents. 
Our model is evaluated in a zero-shot setting, meaning that we use them to predict relevance scores for query-document pairs in languages never seen during training.  
Our results show that the proposed approach can significantly outperform unsupervised retrieval techniques for Arabic, Chinese Mandarin, and Spanish. We also show that augmenting the English training collection with some examples from the target language can sometimes improve performance.

\end{abstract}

\input{1-intro.tex}
\input{2-meth.tex}
\input{3-exp.tex}

\input{4-conc.tex}

\bibliographystyle{splncs04}
\bibliography{biblio}

\end{document}

%% file: 1-intro.tex

\section{Introduction}\label{sec:intro}

Every day, billions of non-English speaking users \cite{owid2019internet} interact with search engines; 
however, commercial retrieval systems have been traditionally tailored to English queries, causing an information access divide between those who can and those who cannot speak this language \cite{young2015digdivide}. 
Non-English search applications have been equally under-studied by most information retrieval researchers. 
Historically, ad-hoc retrieval systems have been primarily designed, trained, and evaluated on English corpora (e.g., \cite{amati2002probabilistic,cao2007learning,carpineto2012survey,metzler2005mdf}).
More recently, a new wave of supervised state-of-the-art ranking models have been proposed by researchers \cite{guo2016deep,hui2017pacrr,macavaney2019cedr,mitra2018introduction,onal2018neural,xiong2017end,yang2019simple}; 
these models rely on neural architectures 
to rerank the head of search results retrieved using a traditional unsupervised ranking algorithm, such as BM25.
Like previous ad-hoc ranking algorithms, these methods are almost exclusively trained and evaluated on English queries and documents.

The absence of rankers designed to operate on languages other than English can largely be attributed to a lack of suitable  publicly available data sets. 
This aspect particularly limits supervised ranking methods, as they require samples for training and validation. 
For English, previous research relied on English collections such as TREC Robust 2004 \cite{voorhees2005overview}, the 2009-2014 TREC Web Track \cite{collins2015trec}, and 
MS MARCO \cite{bajaj2016ms}. 
No datasets of similar size exist for other languages.

While most of recent approaches have focused on ad hoc retrieval for English, some researchers have studied the problem of cross-lingual information retrieval. 
Under this setting, document collections are typically in English, while queries get translated to several languages; sometimes, the opposite setup is used.
Throughout the years, several cross lingual tracks were included as part of TREC. 
TREC 6, 7, 8 \cite{braschler2000cross} offer queries in English, German, Dutch, Spanish, French, and Italian.
For all three years, the document collection was kept in English. 
CLEF also hosted multiple cross-lingual ad-hoc retrieval tasks from 2000 to 2009 \cite{braschler2003clef}. 
Early systems for these tasks leveraged dictionary and statistical translation approaches, as well as other indexing optimizations \cite{peters2012multilingual}. 
More recently, approaches that rely on cross-lingual semantic representations (such as multilingual word embeddings) have been explored. 
For example, Vulic and Moens \cite{vulic2015monolingual} proposed BWESG, an algorithm to learn word embeddings on aligned documents that can be used to calculate document-query similarity.
Sasaki et al \cite{sasaki2018cross} leveraged a data set of Wikipedia pages in 25 languages to train a learning to rank algorithm for Japanese-English and Swahili-English cross-language retrieval. 
Litschko et al \cite{litschko2018unsupervised} proposed an unsupervised framework that relies on aligned word embeddings. 
Ultimately, while related, these approaches are only beneficial to users who can understand documents in two or more languages instead of directly tackling non-English document retrieval. 

A few monolingual ad-hoc data sets exist, but most are too small to train a supervised ranking method.
For example, TREC produced several non-English test collections: Spanish \cite{harman1996overview}, Chinese Mandarin  \cite{voorhees1998sixth}, and Arabic \cite{trec2002arabic}. Other languages were explored, but the document collections are no longer available. 
The CLEF initiative includes some non-English monolingual datasets, though these are primarily focused on European languages~\cite{braschler2003clef}.
Recently, Zheng et al.~\cite{zheng2018sogou} introduced Sogou-QCL, a
large query log dataset in Mandarin.
Such datasets are only available for languages that already have large, established search engines.

Inspired by the success of neural retrieval methods, this work focuses on studying the problem of monolingual ad-hoc retrieval on non English languages using supervised neural approaches. 
In particular, to circumvent the lack of training data, we leverage transfer learning techniques to train Arabic, Mandarin, and Spanish retrieval models using English training data. 
In the past few years, transfer learning between languages has been proven to be a remarkably effective approach for low-resource multilingual tasks (e.g. \cite{johnson2017google,kim2017cross,schuster2019dialog,yang2017transfer}).
Our model leverages a pre-trained multi-language transformer model to obtain an encoding for queries and documents in different languages; 
at train time, this encoding is used to predict relevance of query document pairs in English.
We evaluate our models in a zero-shot setting; that is, we use them to predict relevance scores for query document pairs in languages never seen during training. By leveraging a pre-trained multilingual language model, which can be easily trained from abundant aligned \cite{lample2019cross} or unaligned  \cite{conneau2017word} web text, we achieve competitive retrieval performance without having to rely on language specific relevance judgements.
During the peer review of this article, a preprint~\cite{Shi2019CrossLingualRT} was published with similar observations as ours.
In summary, our contributions are:

\begin{itemize}
\item We study zero shot transfer learning for IR in non-English languages. 
\item We propose a simple yet effective technique that leverages contextualized word embedding as multilingual encoder for query and document terms. Our approach outperforms several baselines on multiple non-English collections.
\item We show that including additional in-language training samples may help further improve ranking performance.
\item We release our code for pre-processing, initial retrieval, training, and evaluation of non-English datasets.\footnote{\url{https://github.com/Georgetown-IR-Lab/multilingual-neural-ir}} We hope that this 
encourages others to consider cross-lingual modeling implications in future work.
\end{itemize}

%% file: 2-meth.tex
\section{Methodology}\label{sec:meth}

\noindent\textbf{Zero-shot Multi-Lingual Ranking.}
Because large-scale relevance judgments are largely absent in languages other than English, we propose a new setting to evaluate learning-to-rank approaches: zero-shot cross-lingual ranking. This setting makes use of relevance data from one language that has a considerable amount of training data (e.g., English) for model training and validation, and applies the trained model to a different language for testing.

More formally, let $\sourcecol$ be a collection of relevance tuples in the source language, and $\targetcol$ be a collection of relevance judgments from another language. Each relevance tuple $\tuple{\query,\doc,\rel}$ consists of a query, document, and relevance score, respectively. In typical evaluation environments, $\sourcecol$ is segmented into multiple splits for training ($\sourcecol_{train}$) and testing ($\sourcecol_{test}$), such that there is no overlap of queries between the two splits. A ranking algorithm is tuned on $\sourcecol_{train}$ to define the ranking function $R_{\sourcecol_{train}}(\query,\doc)\in\mathbb{R}$, which is subsequently tested on $\sourcecol_{test}$. We propose instead tuning a model on all data from the source language (i.e., training $R_{\sourcecol}(\cdot)$), and testing on a collection from the second language ($\targetcol$).

%% file: 3-exp.tex

\vspace{2mm}
\noindent\textbf{Datasets.} We evaluate on monolingual newswire datasets from three languages: Arabic, Mandarin, and Spanish. The Arabic document collection contains $384k$ documents (LDC2001T55), and we use topics/relevance information from the 2001--02 TREC Multilingual track (25 and 50 topics, respectively).
For Mandarin, we use $130k$ news articles from LDC2000T52. Mandarin topics and relevance judgments are utilized from TREC 5 and 6 (26 and 28 topics, respectively). Finally, the Spanish collection contains $58k$ articles from LDC2000T51, and we use topics from TREC 3 and 4 (25 topics each). We use the topics, rather than the query descriptions, in all cases except TREC Spanish 4, in which only descriptions are provided. The topics more closely resemble real user queries than descriptions.\footnote{Some have observed that the context provided by query descriptions are valuable for neural ranking, particularly when using contextualized language models~\cite{Dai2019DeeperTU}.} We test on these collections because they are the only document collections available from TREC at this time.\footnote{\url{https://trec.nist.gov/data/docs_noneng.html}}

We index the text content of each document using a modified version of Anserini with support for the languages we investigate~\cite{Yang2018AnseriniRR}. Specifically, we add Anserini support for Lucene's Arabic and Spanish light stemming and stop word list (via \texttt{SpanishAnalyzer} and \texttt{ArabicAnalyzer}). We treat each character in Mandarin text as a single token.

\vspace{2mm}\noindent\textbf{Modeling.} We explore the following ranking models:
\vspace{-2mm}
\begin{itemize}[leftmargin=*]
\item \textbf{Unsupervised baselines}. We use the Anserini~\cite{Yang2018AnseriniRR} implementation of BM25, RM3 query expansion, and the Sequential Dependency Model (SDM) as unsupervised baselines. In the spirit of the zero-shot setting, we use the default parameters from Anserini (i.e., assuming no data of the target language).
\item \textbf{PACRR}~\cite{hui2017pacrr} models n-gram relationships in the text using learned 2D convolutions and max pooling atop a query-document similarity matrix.
\item \textbf{KNRM}~\cite{xiong2017end} uses learned Gaussian kernel pooling functions over the query-document similarity matrix to rank documents.
\item \textbf{Vanilla BERT}~\cite{macavaney2019cedr} uses the BERT~\cite{devlin2019bert} transformer model, with a dense layer atop the classification token to compute a ranking score. To support multiple languages, we use the \texttt{base-multilingual-cased} pretrained weights. These weights were trained on Wikipedia text from 104 languages.
\end{itemize}
\vspace{-2mm}
We use the embedding layer output from \texttt{base-multilingual-cased} model for PACRR and KNRM. In pilot studies, we investigated using cross-lingual MUSE vectors~\cite{conneau2017word} and the output representations from BERT, but found the BERT embeddings to be more effective.

\vspace{2mm}\noindent\textbf{Experimental Setup.} We train and validate models using TREC Robust 2004 collection~\cite{voorhees2005overview}. TREC Robust 2004 contains 249 topics, $528k$ documents, and $311k$ relevance judgments in English (folds 1-4 from~\cite{huston2014parameters} for training, fold 5 for validation). Thus, the model is only exposed to English text in the training and validation stages (though the embedding and contextualized language models \textit{are} trained on large amounts of unlabeled text in the languages). The validation dataset is used for parameter tuning and for the selection of the optimal training epoch (via nDCG@20). We train using pairwise softmax loss with Adam~\cite{Kingma2015AdamAM}.

We evaluate the performance of the trained models by re-ranking the top 100 documents retrieved with BM25. We report MAP, Precision@20, and nDCG@20 to gauge the overall performance of our approach, and the percentage of judged documents in the top 20 ranked documents (judged@20) to evaluate how suitable the datasets are to approaches that did not contribute to the original judgments.

\input{main_table.tex}

\section{Results}
We present the ranking results in Table~\ref{tab:main_results}. We first point out that there is considerable variability in the performance of the unsupervised baselines; in some cases, RM3 and SDM outperform BM25, whereas in other cases they under-perform. Similarly, the PACRR and KNRM neural models also vary in effectiveness, though more frequently perform much worse than BM25. This makes sense because these models capture matching characteristics that are specific to English. For instance, n-gram patterns captured by PACRR for English do not necessarily transfer well to languages with different constituent order, such as Arabic (VSO instead of SVO). An interesting observation is that the Vanilla BERT model (which recall is only tuned on English text) generally outperforms a variety of approaches across three test languages. This is particularly remarkable because it is a single trained model that is effective across all three languages, without any difference in parameters. The exceptions are the Arabic 2001 dataset, in which it performs only comparably to BM25 and the MAP results for Spanish. For Spanish, RM3 is able to substantially improve recall (as evidenced by MAP), and since Vanilla BERT acts as a re-ranker atop BM25, it is unable to take advantage of this improved recall, despite significantly improving the precision-focused metrics. In all cases, Vanilla BERT exhibits judged@20 above 85\%, indicating that these test collections are still valuable for evaluation.

To test whether a small amount of in-language training data can further improve BERT ranking performance, we conduct an experiment that uses the other collection for each language as additional training data. The in-language samples are interleaved into the English training samples. Results for this few-shot setting are shown in Table~\ref{tab:fewshot}. We find that the added topics for Arabic 2001 (+50) and Spanish 4 (+25) significantly improve the performance. This results in a model significantly better than BM25 for Arabic 2001, which suggests that there may be substantial distributional differences in the English TREC 2004 training and Arabic 2001 test collections. We further back this up by training an ``oracle'' BERT model (training on the test data) for Arabic 2001, which yields a model substantially better (P@20=0.7340, nDCG@20=0.8093, MAP=0.4250).

\input{tab_fewshot.tex}

%% file: main_table.tex
{
\renewcommand{\arraystretch}{1.0}
\setlength\tabcolsep{2.72mm}
\begin{table}
\caption{Zero-shot multi-lingual results for various baseline and neural methods. Significant improvements and reductions in performance compared with BM25 are indicated with $\uparrow$ and $\downarrow$, respectively (paired t-test by query, $p<0.05$). 
}
\label{tab:main_results}
\centering\small
\begin{tabular}{lrrrr}
\toprule
Ranker & P@20 & nDCG@20 & MAP & judged@20 \\
\midrule

\multicolumn{4}{l}{\bf Arabic (TREC 2002) \cite{trec2002arabic}} \\
BM25 & 0.3470 & 0.3863 & 0.2804 & 99.0\% \\
BM25 + RM3 & 0.3320 & 0.3705 & $\downarrow$ 0.2641 & 95.1\% \\
SDM & 0.3380 & 0.3775 & $\downarrow$ 0.2572 & 98.1\% \\
PACRR multilingual & 0.3270 & 0.3499 & $\downarrow$ 0.2517 & 96.4\% \\
KNRM multilingual & 0.3210 & $\downarrow$ 0.3415 & $\downarrow$ 0.2503 & 95.2\% \\
Vanilla BERT multilingual & \bf$\uparrow$ 0.3790 & \bf0.4205 & \bf0.2876 & 97.4\% \\
\midrule
\multicolumn{4}{l}{\bf Arabic (TREC 2001) \cite{trec2002arabic}} \\
BM25 & \bf0.5420 & \bf0.5933 & \bf0.3462 & 97.2\% \\
BM25 + RM3 & $\downarrow$ 0.4700 & 0.5458 & $\downarrow$ 0.2903 & 85.6\% \\
SDM & 0.5140 & 0.5843 & 0.3213 & 96.2\% \\
PACRR multilingual & $\downarrow$ 0.3880 & $\downarrow$ 0.3933 & $\downarrow$ 0.2724 & 90.6\% \\
KNRM multilingual & $\downarrow$ 0.4140 & $\downarrow$ 0.4327 & $\downarrow$ 0.2742 & 91.0\% \\
Vanilla BERT multilingual & 0.5240 & 0.5628 & 0.3432 & 91.0\% \\
\midrule
\multicolumn{4}{l}{\bf Mandarin (TREC 6) \cite{voorhees1998sixth}} \\
BM25 & 0.5962 & 0.6409 & 0.3316 & 89.6\% \\
BM25 + RM3 & $\downarrow$ 0.5019 & $\downarrow$ 0.5571 & 0.2696 & 75.6\% \\
SDM & 0.5942 & 0.6320 & 0.3472 & 92.1\% \\
PACRR multilingual & $\downarrow$ 0.4923 & $\downarrow$ 0.5238 & 0.2856 & 79.0\% \\
KNRM multilingual & $\downarrow$ 0.5308 & $\downarrow$ 0.5497 & $\downarrow$ 0.3107 & 80.8\% \\
Vanilla BERT multilingual & \bf$\uparrow$ 0.6615 & \bf$\uparrow$ 0.6959 & \bf$\uparrow$ 0.3589 & 92.7\% \\
\midrule
\multicolumn{4}{l}{\bf Mandarin (TREC 5) \cite{voorhees1996overview}} \\
BM25 & 0.3893 & 0.4113 & 0.2548 & 85.4\% \\
BM25 + RM3 & $\downarrow$ 0.2768 & $\downarrow$ 0.3021 & $\downarrow$ 0.1698 & 64.6\% \\
SDM & $\uparrow$ 0.4536 & $\uparrow$ 0.4744 & $\uparrow$ 0.2855 & 94.1\% \\
PACRR multilingual & 0.3786 & 0.3998 & 0.2331 & 83.2\% \\
KNRM multilingual & $\downarrow$ 0.3232 & $\downarrow$ 0.3449 & $\downarrow$ 0.2223 & 77.5\% \\
Vanilla BERT multilingual & \bf$\uparrow$ 0.4589 & \bf$\uparrow$ 0.5196 & \bf$\uparrow$ 0.2906 & 92.0\% \\
\midrule
\multicolumn{4}{l}{\bf Spanish (TREC 4) \cite{harman1996overview}} \\
BM25 & 0.3080 & 0.3314 & 0.1459 & 83.8\% \\
BM25 + RM3 & 0.3360 & 0.3358 & \bf$\uparrow$ 0.2024 & 85.2\% \\
SDM & 0.2780 & 0.3061 & 0.1377 & 78.6\% \\
PACRR multilingual & 0.2440 & 0.2494 & 0.1294 & 69.4\% \\
KNRM multilingual & 0.3120 & 0.3402 & 0.1444 & 79.2\% \\
Vanilla BERT multilingual & \bf$\uparrow$ 0.4400 & \bf$\uparrow$ 0.4898 & $\uparrow$ 0.1800 & 85.6\% \\
\midrule
\multicolumn{4}{l}{\bf Spanish (TREC 3) \cite{harman1995overview}} \\
BM25 & 0.5220 & 0.5536 & 0.2420 & 84.8\% \\
BM25 + RM3 & $\uparrow$ 0.6100 & 0.6236 & \bf$\uparrow$ 0.3887 & 93.0\% \\
SDM & 0.4920 & 0.5178 & 0.2258 & 83.8\% \\
PACRR multilingual & $\downarrow$ 0.4140 & $\downarrow$ 0.4092 & 0.2260 & 76.0\% \\
KNRM multilingual & 0.5560 & 0.5700 & 0.2449 & 85.2\% \\
Vanilla BERT multilingual & \bf$\uparrow$ 0.6400 & \bf$\uparrow$ 0.6672 & $\uparrow$ 0.2623 & 90.8\% \\

\bottomrule
\end{tabular}
\end{table}
}

%% file: tab_fewshot.tex
{
\renewcommand{\arraystretch}{0.93}
\begin{table}[t]
\caption{Zero-Shot (ZS) and Few-Shot (FS) comparison for Vanilla BERT (multilingual) on each dataset. Within each metric and dataset, the top result is listed in bold. Significant increases from using FS are indicated with $\uparrow$ (paired t-test, $p<0.05$). 
}
\label{tab:fewshot}
\centering\small\scalebox{0.92}{
\begin{tabular}{lrrrrrr}
\toprule
 & \multicolumn{2}{c}{P@20} & \multicolumn{2}{c}{nDCG@20} & \multicolumn{2}{c}{MAP} \\
 \cmidrule(lr){2-3}\cmidrule(lr){4-5}\cmidrule(lr){6-7}
Dataset & ZS & FS & ZS & FS & ZS & FS \\
\midrule

Arabic 2002 &\bf0.3790 & 0.3690 &\bf0.4205 & 0.3905 &\bf0.2876 & 0.2822 \\
Arabic 2001 & 0.5240 &\bf$\uparrow$ 0.6020 & 0.5628 &\bf$\uparrow$ 0.6405 & 0.3432 &\bf0.3529 \\
Mandarin 6 & 0.6615 &\bf0.6808 & 0.6959 &\bf0.7099 &\bf0.3589 & 0.3537 \\
Mandarin 5 & 0.4589 &\bf0.4643 &\bf0.5196 & 0.5014 &\bf0.2906 & 0.2895 \\
Spanish 4 & 0.4400 &\bf$\uparrow$ 0.5060 & 0.4898 &\bf$\uparrow$ 0.5636 & 0.1800 &\bf$\uparrow$ 0.2020 \\
Spanish 3 & 0.6400 &\bf0.6560 & 0.6672 &\bf0.6825 & 0.2623 &\bf0.2684 \\

\bottomrule
\end{tabular}
}\end{table}
}

%% file: 4-conc.tex
\section{Conclusion}\label{sec:conc}

We introduced a zero-shot multilingual setting for evaluation of neural ranking methods. This is an important setting due to the lack of training data available in many languages. We found that contextualized languages models (namely, BERT) have a big upper-hand, and are generally more suitable for cross-lingual performance than prior models (which may rely more heavily on phenomena exclusive to English). We also found that additional in-language training data may improve the performance, though not necessarily. By releasing our code and models, we hope that cross-lingual evaluation will become more commonplace.